# Pseudocontact shifts and paramagnetic susceptibility in semiempirical and quantum chemistry theories

Giacomo Parigi,[1,2,a)] Ladislav Benda,[3,a)] Enrico Ravera,[1,2] Maurizio Romanelli,[4] Claudio Luchinat[1,2]

[1]Magnetic Resonance Center (CERM) and Interuniversity Consortium for Magnetic Resonance of Metallo Proteins (CIRMMP), Via L. Sacconi 6, 50019 Sesto Fiorentino, Italy

[2]Department of Chemistry "Ugo Schiff", University of Florence, Via della Lastruccia 3, 50019 Sesto Fiorentino, Italy

[3]Centre de RMN à Très Hauts Champs, FRE 2034 CNRS / ENS de Lyon / UCB Lyon 1, 5 rue de la Doua, 69100 Villeurbanne (Lyon), France

[4]Department of Earth Sciences, University of Florence, via G. La Pira 4, 50121 Florence, Italy

Pseudocontact shifts are traditionally described as a function of the anisotropy of the paramagnetic susceptibility tensor, according to the semiempirical theory mainly developed by Kurland and McGarvey (Kurland, R. J. & McGarvey, B. R. *J. Magn. Reson.* **2,** 286–301 (1970)). The paramagnetic susceptibility tensor is required to be symmetric. Applying point-dipole approximation to the quantum chemistry theory of hyperfine shift, pseudocontact shifts are found to scale with a non-symmetric tensor that differs by a factor $\mathbf{g}^T/g_e$ from the paramagnetic susceptibility tensor derived within the semiempirical framework. We analyze the foundations of the Kurland–McGarvey pseudocontact shift expression and recall that it is inherently based on the Russell–Saunders (*LS*) coupling approximation for the spin-orbit coupling. We show that the difference between the semiempirical and quantum chemistry pseudocontact shift expressions arises directly from the different treatment of the orbital contribution to the hyperfine coupling.

[a)] Authors to whom correspondence should be addressed. Electronic mail: parigi@cerm.unifi.it, ladislav.benda@ens-lyon.fr

**Introduction**

Paramagnetic systems are attracting more and more attention because the information content of paramagnetic NMR (PNMR) data on molecular structures and dynamics is very valuable, especially in large complexes and in the presence of conformational heterogeneity. Paramagnetic data in fact provide information on the structure of a paramagnetic molecule as the result of the coupling between nuclear and electron magnetic moments, called hyperfine coupling. The equations describing this effect have been developed since the 50s.[1–3] In the last decades considerable efforts have been devoted to the theoretical description of the hyperfine coupling using quantum chemistry approaches.[4] These approaches have been quite successful in the analysis of EPR spectra.[5] Recently, they were thus ported also for the analysis of the paramagnetic NMR observables.[6–11]

The contribution to the NMR shift ascribed to the presence of a paramagnetic center is called the hyperfine shift. The leading-order terms of the hyperfine shift are the contact and pseudocontact shifts. Pseudocontact shift corresponds to the rotational average of the dipole-dipole interaction between nuclear and electron magnetic moment. In the non-relativistic approximation, the electron magnetic moment is equal to that of a bare electron and under isotropic molecular reorientation the dipolar hyperfine interaction is averaged to zero. When the unpaired electrons reside in a molecule, a non-fully quenched orbital magnetic moment is also present, which requires the consideration of spin-orbit coupling, giving rise to pseudocontact shift as well as $g$-shift, $g$-anisotropy, and zero-field splitting (ZFS) observed in EPR.

Following Kurland and McGarvey,[3] pseudocontact shifts are described in the point-dipole approximation through the paramagnetic susceptibility anisotropy tensor, **Δχ**, and depend on the coordinates of the nuclei in the frame of this tensor. Hereafter, we will refer to this theory as the semi-empirical quantum mechanical (SE) framework traditionally used in experimental PNMR. Pseudocontact shifts have been satisfactorily used as restraints in molecular structure calculations and for monitoring conformational heterogeneities.[12–22] This is done by iterative procedures allowing for the recovery of both the **Δχ** tensor and the molecular structure information, taking advantage of the geometrical constraints of chemical bonds and of other structural information from different sources.

Pseudocontact shift can be calculated using quantum chemistry (QC) approaches as a part of total hyperfine shift.[7,10,23–30] The calculations are typically done within the effective spin Hamiltonian framework, requiring the evaluation of the **g** tensor, of the ZFS tensor and of the hyperfine coupling tensor. It was recently shown[31] that a tensor, indicated as $\boldsymbol{\chi}'$, can be associated to the pseudocontact shifts and expressed in the point-dipole approximation as a function of the EPR **g** and ZFS tensors, all calculated using quantum chemistry methods, *i.e.* without the need to make assumptions about the symmetry of the tensors or their orientation in the molecular frame. Since the magnetic susceptibility tensor is fundamentally required to be symmetric, $\boldsymbol{\chi}'$ cannot be the susceptibility tensor but rather an *ad hoc* tensor defined to map the pseudocontact shifts into the molecular structure. The possibility to evaluate this tensor from first principles of quantum mechanics for any given molecular structure represents an attractive perspective because it could assist in the first steps of molecular structure calculation protocols when restraints of different types are not available, and could even open the route for a structure refinement around the paramagnetic center. Obtaining structural information on the coordination environment of a paramagnetic metal ion would be quite important because, in most cases, ligand nuclei cannot be directly observed due to the paramagnetic line broadening which prevents the detection of their NMR signals.

Here we examine and compare the two existing hyperfine shift theory frameworks along with their underlying assumptions and approximations. To keep the comparisons simple and transparent, we frame the entire discussion in terms of the standard EPR spin Hamiltonian valid for many transition metal complexes where only the ground-state multiplet is thermally populated. We give expressions for Curie spin and paramagnetic susceptibility tensor in terms of magnetic property tensors defined by the EPR spin Hamiltonian. Within the SE approach of McConnell-Robertson[1] and Kurland–McGarvey[3] we identify steps in the derivation where the theory relies on the Russell–Saunders (*LS*) coupling approximation. We find that the difference in the pseudocontact shift formula between the SE and QC approaches originates from including the orbital contribution to hyperfine coupling in the SE approach in the form dictated by the *LS* coupling scheme.

**Curie spin and paramagnetic susceptibility**

In the description of a paramagnetic system composed of, e.g., a paramagnetic metal ion and a nucleus, whose NMR signal is observed, we consider the total spin **S** of the unpaired electrons of the metal and the spin **I** of the nucleus. The corresponding magnetic moments are

$$\boldsymbol{\mu}_S = -\mu_B g_e \mathbf{S}, \qquad \boldsymbol{\mu}_I = \hbar \gamma_I \mathbf{I} \qquad (1)$$

We must also consider the presence of electron orbital angular momenta and the existence of spin-orbit coupling (Fig. 1). When all orbital excited states are sufficiently far in energy from the ground state with respect to the energy of the thermal bath, the spin Hamiltonian formalism can be used and the total electron magnetic moment **μ** can be written in the form

$$\boldsymbol{\mu} = -\mu_B \mathbf{g} \cdot \mathbf{S} \qquad (2)$$

with the **g** tensor associated with the ground-state multiplet taking into account the spin-orbit coupling.

Since electrons relax much more efficiently than nuclei, they change their spin state much faster than nuclei. As a result, the unpaired electron(s) change rapidly their state among the possible $m_S$ levels before the nuclei change their own $m_I$ energy levels. This implies that the NMR shift depends on the interaction of the nuclear magnetic moment with a thermally averaged electron magnetic moment, calculated from the population distribution over energy levels at a given temperature. The electron spin **S** can thus be conveniently separated into two terms, one with null thermal average, **s** ($\langle \mathbf{s} \rangle = 0$), and the other, the Curie spin $\mathbf{S}_C$, with thermal average equal to that of **S** ($\langle \mathbf{S}_C \rangle = \langle \mathbf{S} \rangle$):

$$\mathbf{S} = \mathbf{s} + \mathbf{S}_C \qquad (3)$$

In the presence of a magnetic field $\mathbf{B}_0$, the average electron-spin-only magnetic moment is (see Fig. 2)

$$\langle \boldsymbol{\mu}_S \rangle = -\mu_B g_e \langle \mathbf{S}_C \rangle \tag{4}$$

and, when the spin Hamiltonian formalism can be used, the average effective electron magnetic moment can be written as

$$\langle \boldsymbol{\mu} \rangle = -\mu_B \mathbf{g} \cdot \langle \mathbf{S}_C \rangle . \tag{5}$$

Following van Vleck[32] it can be demonstrated (see Appendix A) that in the spin Hamiltonian formalism

$$\langle \mathbf{S}_C \rangle = -\frac{\mu_B}{kT} \langle \mathbf{SS}^T \rangle \cdot \mathbf{g}^T \cdot \mathbf{B}_0 \tag{6}$$

where $^T$ denotes a matrix transpose and $\langle \mathbf{SS}^T \rangle$ is the electron spin dyadic equal to

$$\langle S_u S_v \rangle = \frac{\sum_{mn} Q_{mn} \langle n|S_u|m\rangle\langle m|S_v|n\rangle}{\sum_n \exp[-E_n^0/(kT)]}, \qquad u, v = \{x, y, z\} \tag{7}$$

$$Q_{mn} = \begin{cases} \exp[-E_n^0/(kT)] & \text{for } E_n^0 = E_m^0 \\ -\frac{kT}{E_m^0 - E_n^0}\{\exp[-E_m^0/(kT)] - \exp[-E_n^0/(kT)]\} & \text{for } E_n^0 \neq E_m^0 \end{cases}$$

where $E_n^0$ is the energy of the state $|n\rangle$ at zero magnetic field, and thus depends on the ligand field and the presence of ZFS. From Eqs. (5) and (6),

$$\langle \boldsymbol{\mu} \rangle = \frac{\mu_B^2}{kT} \mathbf{g} \cdot \langle \mathbf{SS}^T \rangle \cdot \mathbf{g}^T \cdot \mathbf{B}_0 . \tag{8}$$

Outside the spin Hamiltonian approximation, the average effective electron magnetic moment is given by (see Appendix A)

$$\langle \boldsymbol{\mu} \rangle = \frac{1}{kT} \frac{\sum_{mn} Q_{mn} \langle n|\boldsymbol{\mu}|m\rangle\langle m|\boldsymbol{\mu}^T|n\rangle}{\sum_n \exp[-E_n^0/(kT)]} \cdot \mathbf{B}_0 . \tag{9}$$

Magnetic susceptibility is defined as the derivative of the magnetization of a substance with respect to the magnetic field strength; if the magnetic field is weak enough not to reach the saturation conditions, the magnetization increases linearly with $\mathbf{B}_0$, and the magnetic susceptibility is thus independent of the magnetic field strength. The paramagnetic susceptibility $\chi$ is thus defined as

$$\chi_{uv} = \mu_0 \left.\frac{d\langle\mu_u\rangle}{dB_{0,v}}\right|_{\mathbf{B_0}=\mathbf{0}}, \qquad u,v = \{x,y,z\} \tag{10}$$

or

$$\langle\boldsymbol{\mu}\rangle = \frac{\boldsymbol{\chi}\cdot\mathbf{B_0}}{\mu_0}. \tag{11}$$

Comparing Eqs. (8) and (11) we have

$$\boldsymbol{\chi} = \frac{\mu_0 \mu_B^2}{kT}\mathbf{g}\cdot\langle\mathbf{SS}^T\rangle\cdot\mathbf{g}^T. \tag{12}$$

The possibility to derive this susceptibility formula was suggested previously.[9] The resulting susceptibility tensor is symmetric, as it should be, given that

$$\boldsymbol{\chi} = -\mu_0 \left.\frac{d^2 E}{d\mathbf{B}_0^T d\mathbf{B}_0}\right|_{\mathbf{B_0}=\mathbf{0}} \tag{13}$$

When contributions from thermally accessible orbital excited states must be included, the general van Vleck expression[32] for the susceptibility tensor should be used

$$\boldsymbol{\chi} = \frac{\mu_0}{kT}\frac{\sum_{mn} Q_{mn}\langle n|\boldsymbol{\mu}|m\rangle\langle m|\boldsymbol{\mu}^T|n\rangle}{\sum_n \exp[-E_n^0/(kT)]}. \tag{14}$$

**Hyperfine shift – the QC view**

In the QC approach, hyperfine shift is the temperature-dependent part of the NMR shift due to the magnetic interactions between nuclei and unpaired electrons. To avoid confusion, it should be emphasized that the direct interaction between the nuclear spin and the electron orbital angular momenta is causing the temperature-independent orbital (Ramsay) shift, *i.e.* the open-shell analogue of the diamagnetic shift. In the analysis of measured hyperfine shifts, it is usually assumed that this contribution is equal to the orbital shift of a diamagnetic analogue of the paramagnetic system. The orbital shift is not to be confused with the effect of temperature-independent paramagnetism which becomes important in the presence of low-lying excited states. In the QC theory, the contribution to the hyperfine shift from the orbital angular momenta of electrons arises exclusively via the spin-orbit coupling.

There are currently two valid formulations of the QC theory of hyperfine shift. One relies on expressing the shielding tensor as the second derivative of the thermally averaged Helmholtz free energy $F$ with respect to the magnetic field and the nuclear magnetic moment, calculated at zero magnetic field and zero magnetic moment.[8,9] The other, adopted also in this work, follows the approach of Kurland–McGarvey, thermally averaging the density matrix operator.[10] In the limit of a single populated orbital ground state the standard EPR spin Hamiltonian can be invoked,

$$H^{\mathrm{epr}} = \mathbf{S}^T \cdot \mathbf{D} \cdot \mathbf{S} + \mu_\mathrm{B} \mathbf{B}_0 \cdot \mathbf{g} \cdot \mathbf{S} + \mathbf{S}^T \cdot \mathbf{A} \cdot \mathbf{I}, \tag{15}$$

where the ZFS, Zeeman, and hyperfine coupling terms of the spin Hamiltonian define the ZFS tensor $\mathbf{D}$, the $\mathbf{g}$ tensor, and the hyperfine coupling tensor $\mathbf{A}$, respectively. The hyperfine shielding tensor $\boldsymbol{\sigma}^{\mathrm{hf}}$ is then expressed as[10]

$$\boldsymbol{\sigma}^{\mathrm{hf}} = -\frac{\mu_\mathrm{B}}{\hbar \gamma_I kT} \mathbf{g} \cdot \langle \mathbf{SS}^T \rangle \cdot \mathbf{A}, \tag{16}$$

where $\langle \mathbf{SS}^T \rangle$ is the electron spin dyadic of Eq. (7).

The leading-order terms of $\mathbf{A}$ are the spin-dipolar and Fermi-contact coupling tensor,

$$\mathbf{A} = \mathbf{A}^{\mathrm{sd}} + \mathbf{A}^{\mathrm{con}} = \mathbf{A}^{\mathrm{sd}} + A^{\mathrm{con}} \mathbf{1} \tag{17}$$

$$\mathbf{A}^{\mathrm{sd}} = \frac{\mu_0}{4\pi} \hbar \gamma_I \mu_\mathrm{B} g_\mathrm{e} \int \frac{3\mathbf{rr}^T - r^2 \mathbf{1}}{r^5} \frac{\rho(\mathbf{r})}{2S} \mathrm{d}^3 r \tag{18}$$

$$A^{\mathrm{con}} = \frac{\mu_0}{3} \hbar \gamma_I \mu_\mathrm{B} g_\mathrm{e} \frac{\rho_I}{S} \tag{19}$$

where $\mathbf{r}$ is the vector between the paramagnetic center and the NMR nucleus, $\mathbf{1}$ is a $3 \times 3$ unit matrix, $\rho(\mathbf{r})$ is the electron spin density distribution, and $\rho_I$ is the contact spin density at the NMR nucleus. For nuclei far from the paramagnetic center, the point-dipole approximation applies and the spin-dipolar hyperfine coupling can be expressed as

$$\mathbf{A}^{\mathrm{sd}} \approx \frac{\mu_0}{4\pi} \frac{\hbar \gamma_I \mu_\mathrm{B}}{r^3} g_\mathrm{e} \left( \frac{3\mathbf{rr}^T}{r^2} - \mathbf{1} \right). \tag{20}$$

The part of hyperfine shielding tensor in Eq. (16) due to $\mathbf{A}^{\mathrm{sd}}$ was in ref. [31] identified with the dipolar shielding tensor

$$\boldsymbol{\sigma}^{\mathrm{dip}} \approx -\frac{\mu_0}{4\pi r^3} \frac{\mu_\mathrm{B}^2}{kT} g_\mathrm{e} \mathbf{g} \cdot \langle \mathbf{SS}^T \rangle \cdot \left( \frac{3\mathbf{rr}^T}{r^2} - \mathbf{1} \right). \tag{21}$$

Defining a non-symmetric tensor

$$\boldsymbol{\chi}' = \frac{\mu_0 \mu_B^2}{kT} g_e \mathbf{g} \cdot \langle \mathbf{SS}^T \rangle, \qquad (22)$$

Eq. (21) can be recast in analogy to the well-established dipolar shielding formula of the SE approach (see Eq. (34) below) as

$$\boldsymbol{\sigma}^{\mathrm{dip}} \approx -\frac{1}{4\pi r^3} \boldsymbol{\chi}' \cdot \left( \frac{3\mathbf{rr}^T}{r^2} - \mathbf{1} \right). \qquad (23)$$

In the QC approach $\boldsymbol{\chi}'$ assumes the place traditionally reserved for $\boldsymbol{\chi}$. In essence, the outlined QC derivation within the point-dipole approximation provides a dipolar shielding expression that differs by a missing factor $\mathbf{g}^T/g_e$ from the SE approach (see below).

When spin-orbit coupling is fully considered in the QC framework up to the leading order, an additional term $\mathbf{A}^{\mathrm{so}}$ arises in Eq. (17) together with an associated contribution to the hyperfine shielding in Eq. (16). The form of $\mathbf{A}^{\mathrm{so}}$ involves derivatives of orbital hyperfine (or also paramagnetic-spin-orbit, PSO) and spin-orbit (SO) interaction terms $H^{\mathrm{pso}}$ and $H^{\mathrm{so}}$ of the Breit-Pauli quantum-mechanical Hamiltonian (see Fig. 1). $\mathbf{A}^{\mathrm{so}}$ is thus very different from both $\mathbf{A}^{\mathrm{con}}$ and $\mathbf{A}^{\mathrm{sd}}$:[5,33]

$$\mathbf{A}^{\mathrm{so}} = \sum_m \frac{\left\langle 0 \left| \frac{\partial H^{\mathrm{pso}}}{\partial \mathbf{I}} \right| m \right\rangle \left\langle m \left| \sum_i \frac{\partial H^{\mathrm{so}}}{\partial \mathbf{s}_i^T} \right| 0 \right\rangle + \left\langle 0 \left| \sum_i \frac{\partial H^{\mathrm{so}}}{\partial \mathbf{s}_i} \right| m \right\rangle \left\langle m \left| \frac{\partial H^{\mathrm{pso}}}{\partial \mathbf{I}^T} \right| 0 \right\rangle}{E_0^0 - E_m^0} \qquad (24)$$

where $\mathbf{s}_i$ is the spin of $i$th electron, $H^{\mathrm{pso}}$ has the form

$$H^{\mathrm{pso}} = 2 \frac{\mu_0}{4\pi} \hbar \gamma_I \mu_B \sum_i \frac{\mathbf{l}_{iI}^T}{r_{iI}^3} \cdot \mathbf{I}, \qquad (25)$$

and $\mathbf{l}_{iI}$ is the orbital angular momentum of $i$th electron at the position $\mathbf{r}_{iI}$ relative to the NMR nucleus $I$. The exact form of the spin-orbit Hamiltonian $H^{\mathrm{so}}$ can be found for example in Refs. [4,33]. Mean-field approximations of $H^{\mathrm{so}}$ are often used for practical QC calculations.[5] The associated hyperfine shielding term

$$\boldsymbol{\sigma}^{\mathrm{so}} = -\frac{\mu_B}{\hbar \gamma_I kT} \mathbf{g} \cdot \langle \mathbf{SS}^T \rangle \cdot \mathbf{A}^{\mathrm{so}} \qquad (26)$$

thus cannot be simply assigned to either the contact or the pseudocontact part of hyperfine shift. Hence, strictly speaking the comparison of calculated contact and spin-dipolar parts of hyperfine shift to the experimentally distinguished contact and pseudocontact shifts is only meaningful as long as $A^{so}$ is negligible. In the opposite case, only the total hyperfine shift should be compared to the experiment. The non-symmetric $\chi'$ tensor expression (Eq. 22) was derived while neglecting $A^{so}$ altogether.

In Appendix B we show that it is possible to obtain the hyperfine shielding tensor equation (16) directly from its definition as the total second derivative of thermally averaged electronic energy of a paramagnetic system in a magnetic field taken at at zero magnetic field and zero nuclear magnetic moment,

$$\boldsymbol{\sigma}^{\text{hf}} = \frac{d^2 \langle E^{\text{para}} \rangle}{d\mathbf{B}_0^T d\boldsymbol{\mu}_I}\bigg|_{\mathbf{B}_0 = \boldsymbol{\mu}_I = 0}, \tag{27}$$

and applying the van Vleck's perturbational approach, which provides

$$\boldsymbol{\sigma}^{\text{hf}} = -\frac{1}{kT} \frac{\sum_{mn} Q_{mn} \left\langle n \left| \frac{\partial H^{(1)}}{\partial \mathbf{B}_0^T} \right| m \right\rangle \left\langle m \left| \frac{\partial H^{(1)}}{\partial \boldsymbol{\mu}_I} \right| n \right\rangle}{\sum_n \exp(-E_n^0/kT)} \tag{28}$$

where $H^{(1)} = -\mathbf{B}_0^T \cdot \boldsymbol{\mu} + H^{\text{hfc}}$ with $H^{\text{hfc}}$ defined as the third term of Eq. (15).

**Hyperfine shift – the SE view**

The traditional SE theory of hyperfine shift is built on the assumption of the validity of the *LS* coupling scheme.[3,34] In this approximate spin-orbit coupling regime, which strictly applies only to light atoms, the total orbital angular momentum $\hat{\mathbf{L}}$ is a Hermitian operator with an associated observable $\mathbf{L}$ such that

$$H^{\text{so}} \approx H^{LS} = \lambda \mathbf{S}^T \cdot \mathbf{L} \tag{29}$$

The orbital magnetic moment of a paramagnetic center is then

$$\boldsymbol{\mu}_L = -\mu_B \mathbf{L} \tag{30}$$

and the total electron magnetic moment adopts the form

$$\boldsymbol{\mu} \approx \boldsymbol{\mu}_{S+L} = -\mu_B (\mathbf{L} + g_e \mathbf{S}). \tag{31}$$

NMR shift is the relative difference in the resonance frequency of a nuclear transition between states differing by $\Delta m_I = \pm 1$ in the absence of the paramagnetic center as well as in its presence. In the SE theory developed by McConnell–Robertson[1] and Kurland–McGarvey[3], the hyperfine shift is related to the energy of the hyperfine coupling $E^{\text{hfc}}$ corresponding to the interactions of the nuclear magnetic moment with the total average magnetic moment of the metal center, i.e., with $\langle \boldsymbol{\mu}_{S+L} \rangle$.

If the shift is expressed in ppm, the hyperfine coupling energy $E^{\text{hfc}}$ should be scaled by the nuclear Larmor frequency ($\gamma_I B_0 / 2\pi$). Therefore, the hyperfine shift is[3,35]

$$\delta^{\text{hf}} = -\frac{E^{\text{hfc}}}{\hbar \gamma_I B_0 m_I}. \tag{32}$$

Two leading-order terms of the hyperfine coupling are usually distinguished, the long-range through-space dipolar interaction and the short-range through-bond contact term. In the long-range limit the point-dipole approximation applies and the energy associated with the dipole-dipole interaction between nuclear magnetic moment and effective electron magnetic moment is[1]

$$E^{\text{dip}} = -\frac{\mu_0}{4\pi r^3} \langle \boldsymbol{\mu}_{S+L} \rangle^T \cdot \left( \frac{3\mathbf{r}\mathbf{r}^T}{r^2} - \mathbf{1} \right) \cdot \boldsymbol{\mu}_I. \tag{33}$$

Here we used the fact that in the $LS$ coupling regime the interaction between the orbital magnetic moment $\boldsymbol{\mu}_L$ and the nuclear magnetic moment $\boldsymbol{\mu}_I$ has the same dipolar form as the interaction between $\boldsymbol{\mu}_S$ and $\boldsymbol{\mu}_I$ (see Eq. (42) and Appendix C).[34,36] Outside of the $LS$ coupling regime, Eq. (33) may not apply, and the PSO Hamiltonian, Eq. (25), and thus the QC approach, could be preferable to describe the interaction between the electron orbital angular momenta and the nuclear spin. The form of the dipolar hyperfine interaction energy in Eq. (33) implies, together with the definition of susceptibility tensor in Eq. (11) and the definition of hyperfine shielding tensor in Eq. (27), the well-known relation between the dipolar shielding tensor and the susceptibility tensor[35]

$$\boldsymbol{\sigma}^{\text{dip}} = -\frac{1}{4\pi r^3} \boldsymbol{\chi} \cdot \left( \frac{3\mathbf{r}\mathbf{r}^T}{r^2} - \mathbf{1} \right), \tag{34}$$

in agreement with Kurland & McGarvey (see Appendix D). Note that the dipolar shielding expression of Eq. (34) is derived independently of the spin Hamiltonian formalism and assumes only the validity of the $LS$ coupling scheme.

In contrast, the contact term is traditionally viewed as the interaction between the nucleus and the part of the electron magnetic moment corresponding to the electron spin density delocalized on the nucleus. The energy associated with the contact interaction is thus

$$E^{\mathrm{con}} = -\frac{\mu_0}{3}\frac{\rho_I}{S}\langle\boldsymbol{\mu}_S\rangle^T \cdot \boldsymbol{\mu}_I \tag{35}$$

where $\rho_I$ is the contact electron spin density at the NMR nucleus. Note that the *LS* coupling scheme is not invoked in this case.

From Eqs. (4)-(6) and (32)-(35) we find the hyperfine shift corresponding to a single orientation of the external magnetic field and of the nuclear magnetic moment relative to the molecule

$$\delta^{\mathrm{hf}} = \boldsymbol{\kappa}^T \cdot \left[\frac{\mu_0 \mu_B^2}{kT}\frac{\mathbf{g}\cdot\langle\mathbf{SS}^T\rangle\cdot\mathbf{g}^T}{4\pi r^3}\cdot\left(\frac{3\mathbf{rr}^T}{r^2}-\mathbf{1}\right) + \frac{\mu_0 \mu_B^2}{3kT}\frac{\rho_I}{S}g_e\mathbf{g}\cdot\langle\mathbf{SS}^T\rangle\right]\cdot\boldsymbol{\iota} \tag{36}$$

where $\boldsymbol{\kappa}$ and $\boldsymbol{\iota}$ are a unit vectors in the magnetic field direction and in the direction of the quantization axis of **I**, respectively. The hyperfine shielding tensor is thus expressed as

$$\boldsymbol{\sigma}^{\mathrm{hf}} = -\frac{\mu_0 \mu_B^2}{kT}\mathbf{g}\cdot\langle\mathbf{SS}^T\rangle\cdot\left[\frac{1}{4\pi r^3}\mathbf{g}^T\cdot\left(\frac{3\mathbf{rr}^T}{r^2}-\mathbf{1}\right) + \frac{1}{3}\frac{\rho_I}{S}g_e\mathbf{1}\right]. \tag{37}$$

Outside the validity of the spin Hamiltonian formalism, Eq. (9) should rather be used instead of Eq. (8) to take into account the contributions from orbital excited states. Furthermore, the second term of Eq. (36) describing the contact shift assumes that the contact spin density $\rho_I$ is constant over all states: if orbital excited states should be considered, each of them would contribute with a different, state-specific contact spin density, and the Kurland–McGarvey equation for contact shift would be recovered.

As clear from Eq. (36), in the high-field case when the nuclear spin is oriented along the magnetic field, $\boldsymbol{\sigma}^{\mathrm{hf}}$ allows for an easy calculation of the hyperfine shift for different directions of the magnetic field. In the case of isotropic molecular reorientation, the hyperfine shift becomes

$$\delta^{\mathrm{hf}} = -\frac{1}{3}\mathrm{Tr}(\boldsymbol{\sigma}^{\mathrm{hf}}) \approx -\frac{1}{3}\mathrm{Tr}(\boldsymbol{\sigma}^{\mathrm{con}}) - \frac{1}{3}\mathrm{Tr}(\boldsymbol{\sigma}^{\mathrm{dip}}) = \delta^{\mathrm{con}} + \delta^{\mathrm{pc}} \tag{38}$$

where Tr denotes trace of a matrix, $\delta^{\mathrm{con}}$ is the contact shift, and $\delta^{\mathrm{pc}}$ is the pseudocontact shift.[35]

The hyperfine shielding tensor expression in the point-dipole approximation, Eq. (37), can be formally recast in the form analogous to Eq. (16) of the QC approach[10] as

$$\boldsymbol{\sigma}^{\text{hf}} = -\frac{\mu_B}{\hbar\gamma_I kT} \mathbf{g} \cdot \langle \mathbf{SS}^T \rangle \cdot \mathbf{A} \tag{39}$$

where

$$\mathbf{A} = \mathbf{A}^{\text{dip}} + \mathbf{A}^{\text{con}} = \mathbf{A}^{\text{dip}} + A^{\text{con}}\mathbf{1} \tag{40}$$

$$\mathbf{A}^{\text{dip}} = \mathbf{A}^{\text{sd}} + \mathbf{A}^L = \frac{\mu_0}{4\pi}\frac{\hbar\gamma_I\mu_B}{r^3}\mathbf{g}^T \cdot \left(\frac{3\mathbf{rr}^T}{r^2} - \mathbf{1}\right) \tag{41}$$

$$\mathbf{A}^L = \frac{\mu_0}{4\pi}\frac{\hbar\gamma_I\mu_B}{r^3}(\mathbf{g}^T - g_e\mathbf{1}) \cdot \left(\frac{3\mathbf{rr}^T}{r^2} - \mathbf{1}\right) \tag{42}$$

$$A^{\text{con}} = \frac{\mu_0}{3}\hbar\gamma_I\mu_B g_e \frac{\rho_I}{S} \tag{43}$$

When the orbital contributions to **A** are omitted, $\boldsymbol{\mu}_{S+L}$ in Eq. (33) becomes $\boldsymbol{\mu}_S$, the "orbital" term $\mathbf{A}^L$ of Eq. (41) approximating the effects of the interaction of the electron orbital angular momenta with the nuclear magnetic moment is zero, $\mathbf{A}^{\text{dip}}$ consists of $\mathbf{A}^{\text{sd}}$ only, and we obtain the same equations as in the QC framework within the point-dipole approximation. Therefore, the SE framework differs from the QC approach of Ref. [31] by (approximately) including the "orbital correction" to non-relativistic hyperfine coupling in the form of $\mathbf{A}^L$.

**Discussion**

NMR pseudocontact shifts have been and are still widely analysed using the equations derived in the SE formulation of the theory by McConnell–Robertson and Kurland–McGarvey, but there exists a growing QC-based community where a distinct theory framework is being used.[7,10,23–31,37] We have shown using the spin Hamiltonian formalism that the two frameworks imply two different equations for pseudocontact shift where the former depends on the anisotropy of a symmetric paramagnetic susceptibility tensor $\boldsymbol{\chi}$ and the latter on the anisotropy of a previously introduced[31] non-symmetric $\boldsymbol{\chi}'$ tensor, differing from $\boldsymbol{\chi}$ by a missing factor $\mathbf{g}^T/g_e$ (see Fig. 3).

We have traced the difference between the SE and QC hyperfine shift frameworks back to the underlying hyperfine coupling expressions. In the SE approach, the electron orbital angular momenta contribute to hyperfine coupling via the tensor $\mathbf{A}^L$. In contrast, the QC approach neglects the orbital contribution to hyperfine coupling in the first order[31] and allows it to be included in the leading relativistic order via the tensor $\mathbf{A}^{so}$. It should be emphasized that the "orbital" hyperfine coupling term $\mathbf{A}^L$ in the dipolar form of Eq. (42) does not correspond to any of the hyperfine coupling terms known in standard quantum chemistry.[5] $\mathbf{A}^L$ was originally derived as a contribution to the hyperfine coupling tensor due to the PSO Hamiltonian term within the $LS$ coupling approximation, applying second-order perturbation theory with perturbation Hamiltonians $H^{IL}$ and $H^{LS}$ (see Fig. 1 and Appendix C).[34,36] In the QC approach, second-order perturbation theory is also applied but in this case directly to the exact molecular Hamiltonian, resulting in the linear-response $\mathbf{A}^{so}$ term of Eq. (24) involving the exact perturbation Hamiltonians $H^{pso}$ and $H^{so}$.[5] Hence, our comparison reveals that the difference between the two hyperfine shift theory frameworks originates from different initial considerations, taking the $LS$ coupling approximation in the SE approach and the Breit-Pauli molecular Hamiltonian in the QC approach.

The functional forms of $\mathbf{A}^{so}$ (Eq. (24)) and $\mathbf{A}^L$ (Eq. (42)) are very different. To show whether they provide quantitatively similar values, we have explicitly calculated the isotropic hyperfine shielding due to both $\mathbf{A}^{so}$ and $\mathbf{A}^L$ for the atoms near the metal center of cobalt(II)-substituted matrix metalloproteinase 12 (CoMMP-12), a protein previously studied in Benda et al.[31] (see Supporting Information, Table S1). We found that the (QC) $\mathbf{A}^{so}$ in this system was by an order of magnitude smaller than the (SE) $\mathbf{A}^L$. This was true for most of the atoms investigated. The sign of the isotropic hyperfine shielding due to $\mathbf{A}^{so}$ and $\mathbf{A}^L$ was the same in most cases but there were a few atoms where even the signs were opposite. This raises the question of the validity of the $LS$ coupling approximation in the given Co(II) complex.

Hence, we have shown here that the difference between the two theoretical approaches originates from the description of the interaction between the electron orbital magnetic moment and the nuclear magnetic moment. This interaction is approximated to have the same angular dependence as the spin-dipolar interaction by Kurland-McGarvey, and to provide a much smaller contribution to the hyperfine shift in the QC approach.

The two tensors of Eqs. (12) and (22) can in certain systems provide pseudocontact shift values differing by a factor that can be even larger than 2, as a result of the different anisotropy of the $\boldsymbol{\chi}$ and $\boldsymbol{\chi}'$ tensors calculated from the same **g** and ZFS tensors, depending on the magnitude of *g*-anisotropy. For instance, in the case of CoMMP-12, an axial anisotropy $\Delta\chi_{ax}$ of the paramagnetic susceptibility tensor of $12.5 \times 10^{-32}$ m$^3$ was calculated using Eq. (22) and the protein structure PDB 1RMZ,[31] to be compared with a value of $21.3 \times 10^{-32}$ m$^3$, calculated using Eq. (12) and the same protein structure. The calculated magnitude of *g*-anisotropy was 0.65 in this particular case.

What remains to be clarified is which of the two formulations works better to describe the experimental data. In both formulations the pseudocontact shifts depend on the anisotropy of the **g** tensor and on the ZFS tensor. Therefore, accurate experimental values for all these quantities (pseudocontact shifts, **g** tensor and ZFS tensor) are needed to check their relationship. At the time of the submission of this manuscript, to our knowledge all these data were not available for any system. Nevertheless, in the case of the already mentioned CoMMP-12 protein, pseudocontact shifts were measured for nuclei far enough from the paramagnetic metal to exclude any possible Fermi-contact contribution to the observed NMR shift and to ensure the validity of the point-dipole approximation, and the **g** tensor and the ZFS tensor were calculated[31] with well-established correlated multi-reference methods. In this case, the calculations seem to indicate that the non-symmetric $\boldsymbol{\chi}'$ tensor equation provides pseudocontact shifts in better agreement with the experimental data, from which an axial anisotropy $\Delta\chi_{ax}$ of $10.0 \times 10^{-32}$ m$^3$ was obtained. However, the calculations also showed that very small changes in the position of the ligands around the metal center can change dramatically the calculated **g** and ZFS tensors. Although high spin cobalt(II) might not be an ideal case for the calculation of the pseudocontact shifts within the spin Hamiltonian formalism, because of the possible presence of excited states above the ground-state quartet, the lowest lying excited states were calculated to be above 1000 cm$^{-1}$ for this system, i.e. essentially outside the room temperature thermal bath.[31]

Since the submission of this manuscript, two important papers[37,38] and a review[39] were published dealing with the analysis of the pseudocontact shifts against the **g** and ZFS tensors, confirming the need of clarifying which relationship should be used to obtain the pseudocontact shifts from the EPR tensors. One paper, by Grandinetti and coworkers,[38] shows that the $^2$H paramagnetic shifts measured for first row transition metal (manganese, iron, cobalt, nickel, or copper) chloride dihydrates can be well interpreted using the SE Kurland–McGarvey equation, i.e. as a function of the symmetric magnetic susceptibility anisotropy tensors. The latter were nicely predicted using the generalized Van Vleck equation from the **g** tensor, the ZFS tensor and the temperature independent orbital angular momentum dequenching tensor. The authors note an overall agreement of the calculated shift values from calculated **χ** tensors and the experimental data. They also note that an overestimation of the calculated rhombic anisotropy is present, which could hint to the need of further refining the theoretical treatment.

A second paper, by Mareš and Vaara,[37] presents QC calculations for the prediction of the pseudocontact shifts measured for a cobalt(II) complex with an aliphatic 16-carbon-atom chain attached to it. This system was chosen because it contains atoms far enough from the metal so as paramagnetic shifts are not affected by Fermi-contact contributions and the point-dipole approximation is valid. Contributions due to **A**$^{so}$ were assumed to be negligible. Both the **χ** and the **χ′** tensors were determined from the calculated **g** and ZFS tensors, and the pseudocontact shifts resulted in much better agreement with the **χ′** tensor, rather than with the **χ** tensor. This suggests that the non-symmetric QC equation for the **χ′** tensor is more accurate to predict the pseudocontact shifts. However, we note that the calculated *g*-anisotropy was by a factor of about 5 larger than the experimental one. Experimental values for both the *g*-anisotropy and the ZFS, estimated with SQUID data, are in fact available.[40] If these values were used instead of the calculated ones, the susceptibility tensor fitted to experimental pseudocontact shifts ($\Delta\chi_{ax} = 10 \times 10^{-3}$ m$^3$) would rather be in significantly better agreement with the symmetric $\boldsymbol{\chi}$ tensor ($\Delta\chi_{ax} = 10 \times 10^{-32}$ m$^3$) than with the $\boldsymbol{\chi}'$ tensor ($\Delta\chi_{ax} = 7 \times 10^{-32}$ m$^3$). In other words, the symmetric susceptibility equation would actually seem more accurate to predict the pseudocontact shifts if experimental $g$-anisotropy and ZFS were used in the calculations. We note that the experimentally determined $g$-anisotropy seems exceedingly small with respect to the magnitude of the ZFS, which may point to inaccuracy of at least one of the two experimentally derived EPR parameters. We also note that in this particular system the large mobility of the attached carbon chain may considerably complicate the picture, as the position of each atom is not fixed with respect to any metal-centered tensor, so that the predicted pseudocontact shifts should be averaged over the values obtained for each position in a conformational ensemble. Our conclusion is that, besides the possible presence of first-order orbital contributions, the limited accuracy in the calculation of the **g** and ZFS tensors, also due to the structural noise present in the molecular models, does not allow to unequivocally conclude which of the two equations is more accurate.

Finally, in a recent review[39] Pell and co-workers have derived the expression for paramagnetic susceptibility within the spin Hamiltonian framework, our Eq. (12), and they have suggested a way how to reconcile the inconsistency between the QC and SE frameworks. They proposed that the Kurland–McGarvey formula of the SE approach relating dipolar shielding to paramagnetic susceptibility tensor, here Eq. (34), should be modified to the form

$$\boldsymbol{\sigma}^{\text{dip}} = -\frac{g_e}{4\pi r^3}\boldsymbol{\chi} \cdot [\mathbf{g}^T]^{-1} \cdot \left(\frac{3\mathbf{rr}^T}{r^2} - \mathbf{1}\right). \qquad (44)$$

Given that $\boldsymbol{\chi}' = \boldsymbol{\chi} \cdot [\mathbf{g}^T]^{-1} g_e$, this equation coincides with Eq. (23) of the QC approach. Pell et al. also noted that the $\boldsymbol{\chi} \cdot [\mathbf{g}^T]^{-1}$ factor is already contained in the Kurland–McGarvey formula for the contact shielding.[3] Our analysis performed in this work examines the conditions under which such a modification to the traditional SE formula would be needed. As long as the *LS* coupling scheme is valid, $\mathbf{A}^L$ is a good approximation for the "orbital" hyperfine coupling and the Kurland–McGarvey formula applies. Outside of the *LS* coupling regime, the QC approach might be preferable. On the other hand, when the EPR spin Hamiltonian cannot be used because excited states are thermally populated, as in the case of lanthanoid(III) complexes,[41] a closed relation between the most general QC expression for hyperfine shielding[8,9] and van Vleck's paramagnetic susceptibility is not available and the SE Eq. (34) remains a useful approximation.

**Summary and Conclusion**

The paramagnetic susceptibility tensor has been derived using the van Vleck's perturbational approach employing molecular property tensors. This provides a symmetric susceptibility expression in agreement with the SE approach and consistent with the natural symmetry requirement. In the QC derivation of the nuclear shift induced by the isotropic rotational average of the electron-nucleus dipole-dipole interaction in the point-dipole approximation, the pseudocontact shift expression involves a non-symmetric tensor in place of the paramagnetic susceptibility. We have shown here that this discrepancy arises solely from the difference in describing the effects of the orbital angular momenta of electrons on the hyperfine shift and that the two theoretical frameworks coincide when these contributions are not present. In the SE approach, the "orbital" term of hyperfine coupling is derived within the $LS$ coupling scheme[34,36] to have the same dipolar form as the spin-dipolar term that corresponds to the interaction between nuclear spin and total electron spin. In such a case, both of these terms are found to contribute to the nuclear shift by a similar amount and the shift due to the "orbital" hyperfine coupling is directly proportional to the $g$-anisotropy value. In the QC framework, the orbital angular momenta of electrons manifest themselves in the spin-orbit correction to non-relativistic hyperfine coupling. It is found here on an example of metal core of the CoMMP-12 protein that the QC spin-orbit term of hyperfine coupling does not provide a sizable contribution to the hyperfine shift. Moreover, it is pointed out that the spin-orbit hyperfine coupling term does not have a dipolar character, and thus the associated contribution to the hyperfine shift does not fit the simple paramagnetic susceptibility scheme within the point-dipole approximation.

We conclude that including the "orbital" contribution to hyperfine shift in the simple dipolar form of the SE theory significantly overshoots the nuclear shift due to spin-orbit term of hyperfine coupling of standard QC, while it is consistent with the pseudocontact shift theory within the *LS* coupling approximation where the Kurland–McGarvey formula applies.

What still needs to be validated is whether the experimental data, obtained as the difference between the nuclear chemical shifts in a paramagnetic system and those in a diamagnetic analogue (for nuclei far enough from the metal to exclude possible contact contributions) are in better agreement with the tensor provided by Eq. (12) or (22). This requires very precise experimental NMR data, and the validation should be performed without using theoretical predictions for the **g** and zero-field splitting tensors since the methods currently available are affected by significant uncertainties which may be additionally enhanced by a strong dependence of these properties on fine structural details of the metal coordination sphere.

**Acknowledgements**

We thank Prof. Martin Kaupp for fruitful discussions. Support from the European Research Council under the European Union's Horizon 2020 Research and Innovation Programme (Grant 648974 "P-MEM-NMR") is acknowledged. The support from Fondazione Cassa di Risparmio di Firenze, MIUR PRIN 2012SK7ASN, and Instruct-ERIC, an ESFRI Landmark, supported by national member subscriptions is also acknowledged. Specifically, we thank the Instruct-ERIC Core Centre CERM, Italy.

**Appendix A: The Curie spin**

Magnetic moment can be defined as a derivative of the energy with respect to the magnetic field

$$\boldsymbol{\mu} = -\left.\frac{\mathrm{d}E}{\mathrm{d}\mathbf{B}_0^T}\right|_{\mathbf{B}_0=\mathbf{0}} \qquad \left(\mu_u = -\left.\frac{\mathrm{d}E}{\mathrm{d}B_{0,u}}\right|_{\mathbf{B}_0=\mathbf{0}}, \text{ with } u = \{x, y, z\}\right).$$

The thermally averaged magnetic moment can be expressed as

$$\langle \boldsymbol{\mu} \rangle = -\frac{\sum_\lambda \frac{dE_\lambda}{dB_0^T}\Big|_{B_0=0} \exp[-E_\lambda/(kT)]}{\sum_\lambda \exp[-E_\lambda/(kT)]}$$

where $E_\lambda$ is the energy of the $\lambda$-th state, $k$ is the Boltzmann constant and $T$ is the temperature. Following van Vleck, let us suppose that $E_\lambda$ can be written to a good approximation through perturbation theory limited to second-order correction to the ground state energy

$$E_\lambda = E_\lambda^0 + E_\lambda^{(1)} + E_\lambda^{(2)} = E_\lambda^0 + \langle \lambda|H^{(1)}|\lambda\rangle + \sum_{\eta \neq \lambda}\frac{\left|\langle \lambda|H^{(1)}|\eta\rangle\right|^2}{E_\lambda^0 - E_\eta^0}$$

where $|\lambda\rangle$ are the eigenstates of the unperturbed (magnetic-field- and electron-magnetic-moment-independent) Hamiltonian $H^{(0)}$. In our case $H^{(1)}$ is the Zeeman energy ($H^{(1)} = -\mathbf{B}_0^T \cdot \boldsymbol{\mu}$), so that $E^{(1)}$ and $E^{(2)}$ for a relatively small $B_0$ are much smaller than the eigenvalues of $H^{(0)}$, $E_\lambda^0$, independent of the magnetic field. Therefore,

$$\langle \boldsymbol{\mu} \rangle = -\frac{\sum_\lambda \frac{d\left(E_\lambda^{(1)}+E_\lambda^{(2)}\right)}{dB_0^T}\Big|_{B_0=0} \exp[-E_\lambda/(kT)]}{\sum_\lambda \exp[-E_\lambda/(kT)]} = \frac{\sum_\lambda \frac{d\left(\langle \lambda|\mathbf{B}_0^T \cdot \boldsymbol{\mu}|\lambda\rangle - \sum_{\eta \neq \lambda}\frac{\left|\langle \lambda|\mathbf{B}_0^T \cdot \boldsymbol{\mu}|\eta\rangle\right|^2}{E_\lambda^0 - E_\eta^0}\right)}{dB_0^T} \exp[-E_\lambda/(kT)]}{\sum_\lambda \exp[-E_\lambda/(kT)]}.$$

Then, to the first order in $E_\lambda/(kT)$,

$$\langle \boldsymbol{\mu} \rangle = \frac{\sum_\lambda \left(\langle \lambda|\boldsymbol{\mu}|\lambda\rangle - 2\sum_{\eta \neq \lambda}\frac{\langle \lambda|\boldsymbol{\mu}|\eta\rangle\langle \eta|\boldsymbol{\mu}^T \cdot \mathbf{B}_0|\lambda\rangle}{E_\lambda^0 - E_\eta^0}\right)\exp[-E_\lambda^0/(kT)]\left[1 - E_\lambda^{(1)}/(kT)\right]}{\sum_\lambda \exp[-E_\lambda^0/(kT)]\left[1 - E_\lambda^{(1)}/(kT)\right]}$$

and retaining only the terms linear in $B_0$

$$\langle \boldsymbol{\mu} \rangle = \frac{\sum_\lambda \left(\frac{\langle \lambda|\boldsymbol{\mu}|\lambda\rangle\langle \lambda|\boldsymbol{\mu}^T|\lambda\rangle}{kT}\exp[-E_\lambda^0/(kT)] - \sum_{\eta \neq \lambda}\frac{\langle \lambda|\boldsymbol{\mu}|\eta\rangle\langle \eta|\boldsymbol{\mu}^T|\lambda\rangle}{E_\lambda^0 - E_\eta^0}\{\exp[-E_\lambda^0/(kT)] - \exp[-E_\eta^0/(kT)]\}\right)}{\sum_\lambda \exp[-E_\lambda^0/(kT)]} \cdot \mathbf{B}_0 \ .$$

This equation is derived under the assumption that all states at zero field are not degenerate. In the presence of states with degeneracy (indicated with the indexes $p$ and $q$), and using the spin Hamiltonian formalism (Eq. (2))

$$\langle \boldsymbol{\mu} \rangle = \frac{\mu_B^2}{kT} \frac{\mathbf{g} \cdot \Sigma_\lambda \left( \Sigma_{pq} \langle \lambda, p|\mathbf{S}|\lambda, q\rangle\langle \lambda, q|\mathbf{S}^T|\lambda, p\rangle \exp[-E_\lambda^0/(kT)] - kT \Sigma_{\eta \neq \lambda} \frac{\Sigma_{pq} \langle \lambda, p|\mathbf{S}|\eta, q\rangle\langle \eta, q|\mathbf{S}^T|\lambda, p\rangle\{\exp[-E_\lambda^0/(kT)] - \exp[-E_\eta^0/(kT)]\}}{E_\lambda^0 - E_\eta^0} \right) \cdot \mathbf{g}^T}{\Sigma_\lambda \exp[-E_\lambda^0/(kT)]} \cdot \mathbf{B}_0$$

which can be written as

$$\langle \boldsymbol{\mu} \rangle = \frac{\mu_B^2}{kT} \mathbf{g} \cdot \langle \mathbf{SS}^T \rangle \cdot \mathbf{g}^T \cdot \mathbf{B}_0$$

and thus (see Eq. (5))

$$\langle \mathbf{S}_C \rangle = -\frac{\mu_B}{kT} \langle \mathbf{SS}^T \rangle \cdot \mathbf{g}^T \cdot \mathbf{B}_0$$

where $\langle \mathbf{SS}^T \rangle$ is the effective electron spin dyadic equal to

$$\langle S_i S_j \rangle = \frac{\Sigma_{mn} Q_{mn} \langle n|S_i|m\rangle\langle m|S_j|n\rangle}{\Sigma_n \exp[-E_n^0/(kT)]}, \qquad i,j = \{x,y,z\}$$

$$Q_{mn} = \begin{cases} \exp[-E_n^0/(kT)] & \text{for } E_n^0 = E_m^0 \\ -\frac{kT}{E_m^0 - E_n^0}\{\exp[-E_m^0/(kT)] - \exp[-E_n^0/(kT)]\} & \text{for } E_n^0 \neq E_m^0 \end{cases}$$

where $E_n^0$ is the energy of the state $|n\rangle$ at zero magnetic field.

**Appendix B: The paramagnetic shielding tensor**

Expanding the thermally averaged total energy $\langle E^{\text{para}} \rangle$ in the paramagnetic shielding tensor definition, Eq. (27), we obtain

$$\boldsymbol{\sigma}^{\text{hf}} = \frac{d^2}{d\mathbf{B}_0^T d\boldsymbol{\mu}_I} \left[ \frac{\Sigma_\lambda E_\lambda \exp[-E_{\text{el},\lambda}/(kT)]}{\Sigma_\lambda \exp[-E_{\text{el},\lambda}/(kT)]} \right]\bigg|_{\mathbf{B}_0 = \boldsymbol{\mu}_I = 0}$$

where $E_\lambda$ is the energy of the electron-nucleus system in the $\lambda$-th state, $E_{\text{el},\lambda}$ is the energy of the nuclear-spin-free system in the $\lambda$-th state, $k$ is the Boltzmann constant and $T$ is the temperature. Following van Vleck, let us suppose that $E_\lambda$ can be written to a good approximation through perturbation theory limited to second-order correction to the ground state energy

$$E_\lambda = E_\lambda^0 + E_\lambda^{(1)} + E_\lambda^{(2)} = E_\lambda^0 + \langle \lambda|H^{(1)}|\lambda\rangle + \Sigma_{\eta \neq \lambda} \frac{|\langle \lambda|H^{(1)}|\eta\rangle|^2}{E_\lambda^0 - E_\eta^0}.$$

In our case $H^{(1)} = -\mathbf{B}_0^T \cdot \boldsymbol{\mu} + H^{\text{hfc}}$, so that $E^{(1)}$ and $E^{(2)}$ are much smaller than the eigenvalues of $H^{(0)}$, $E_\lambda^0$, independently of the magnetic field. Therefore,

$$\boldsymbol{\sigma}^{\text{hf}} = \frac{d^2}{d\mathbf{B}_0^T d\boldsymbol{\mu}_I} \left[ \frac{\sum_\lambda \left( \langle\lambda|H^{(1)}|\lambda\rangle + \sum_{\eta\neq\lambda} \frac{|\langle\lambda|H^{(1)}|\eta\rangle|^2}{E_\lambda^0 - E_\eta^0} \right) \exp[-E_\lambda^0/(kT)] \left[1 - \frac{\langle\lambda|H_{\text{el}}^{(1)}|\lambda\rangle}{kT}\right]}{\sum_\lambda \exp[-E_\lambda^0/(kT)] \left[1 - \frac{\langle\lambda|H_{\text{el}}^{(1)}|\lambda\rangle}{kT}\right]} \right]_{\mathbf{B}_0=\boldsymbol{\mu}_I=0}$$

with $H_{\text{el}}^{(1)} = -\mathbf{B}_0^T \cdot \boldsymbol{\mu}$ and, to the first order in $E_\lambda/(kT)$,

$$\boldsymbol{\sigma}^{\text{hf}} = \frac{d^2}{d\mathbf{B}_0^T d\boldsymbol{\mu}_I} \left[ \frac{\sum_\lambda \left( \langle\lambda|H^{(1)}|\lambda\rangle + \sum_{\eta\neq\lambda} \frac{|\langle\lambda|H^{(1)}|\eta\rangle|^2}{E_\lambda^0 - E_\eta^0} - \langle\lambda|H^{(1)}|\lambda\rangle \frac{\langle\lambda|H_{\text{el}}^{(1)}|\lambda\rangle}{kT} \right) \exp[-E_\lambda^0/(kT)]}{\sum_\lambda \exp[-E_\lambda^0/(kT)]} \right]_{\mathbf{B}_0=\boldsymbol{\mu}_I=0}.$$

Since $\frac{\partial^2 H^{(1)}}{\partial \mathbf{B}_0^T \partial \boldsymbol{\mu}_I} = 0$ (it is equal to the elements of the orbital (Ramsey) shielding tensor if the nuclear Zeeman interaction is included in $H^{(1)}$),

$$\boldsymbol{\sigma}^{\text{hf}} = -\frac{1}{kT} \frac{\sum_\lambda \left( \frac{d^2(\langle\lambda|H^{(1)}|\lambda\rangle\langle\lambda|H_{\text{el}}^{(1)}|\lambda\rangle)}{d\mathbf{B}_0^T d\boldsymbol{\mu}_I} \bigg|_{\mathbf{B}_0=\boldsymbol{\mu}_I=0} - \sum_{\eta\neq\lambda} \frac{kT}{E_\lambda^0 - E_\eta^0} \frac{d^2 |\langle\lambda|H^{(1)}|\eta\rangle|^2}{d\mathbf{B}_0^T d\boldsymbol{\mu}_I} \bigg|_{\mathbf{B}_0=\boldsymbol{\mu}_I=0} \right) \exp[-E_\lambda^0/(kT)]}{\sum_\lambda \exp[-E_\lambda^0/(kT)]}$$

and thus, since $\frac{\partial H_{\text{el}}^{(1)}}{\partial \boldsymbol{\mu}_I} = \mathbf{0}$ and $|\lambda\rangle, |\eta\rangle$ are the unperturbed states,

$$\boldsymbol{\sigma}^{\text{hf}} = -\frac{1}{kT} \frac{\sum_\lambda \left( \left\langle\lambda\left|\frac{\partial H_{\text{el}}^{(1)}}{\partial \mathbf{B}_0^T}\right|\lambda\right\rangle \left\langle\lambda\left|\frac{\partial H^{(1)}}{\partial \boldsymbol{\mu}_I}\right|\lambda\right\rangle - 2\sum_{\eta\neq\lambda} \frac{kT}{E_\lambda^0 - E_\eta^0} \left\langle\lambda\left|\frac{\partial H^{(1)}}{\partial \mathbf{B}_0^T}\right|\eta\right\rangle \left\langle\eta\left|\frac{\partial H^{(1)}}{\partial \boldsymbol{\mu}_I}\right|\lambda\right\rangle \right) \exp[-E_\lambda^0/(kT)]}{\sum_\lambda \exp[-E_\lambda^0/(kT)]}.$$

Noting that $\frac{\partial H_{\text{el}}^{(1)}}{\partial \mathbf{B}_0^T} = \frac{\partial H^{(1)}}{\partial \mathbf{B}_0^T}$ and

$$\frac{\langle\lambda|H^{(1)}|\eta\rangle\langle\eta|H^{(1)}|\lambda\rangle}{E_\lambda^0 - E_\eta^0} \exp[-E_\lambda^0/(kT)] + \frac{\langle\eta|H^{(1)}|\lambda\rangle\langle\lambda|H^{(1)}|\eta\rangle}{E_\eta^0 - E_\lambda^0} \exp[-E_\eta^0/(kT)] =$$

$$= \langle\lambda|H^{(1)}|\eta\rangle\langle\eta|H^{(1)}|\lambda\rangle \frac{\exp[-E_\lambda^0/(kT)] - \exp[-E_\eta^0/(kT)]}{E_\lambda^0 - E_\eta^0},$$

we obtain

$$\boldsymbol{\sigma}^{hf} = -\frac{1}{kT \sum_\lambda \exp[-E_\lambda^0/(kT)]} \sum_\lambda \left( \left\langle \lambda \left| \frac{\partial H^{(1)}}{\partial \mathbf{B}_0^T} \right| \lambda \right\rangle \left\langle \lambda \left| \frac{\partial H^{(1)}}{\partial \boldsymbol{\mu}_I} \right| \lambda \right\rangle \exp[-E_\lambda^0/(kT)] - \right.$$

$$\left. - \sum_{\eta \neq \lambda} \frac{kT}{E_\lambda^0 - E_\eta^0} \left\langle \lambda \left| \frac{\partial H^{(1)}}{\partial \mathbf{B}_0^T} \right| \eta \right\rangle \left\langle \eta \left| \frac{\partial H^{(1)}}{\partial \boldsymbol{\mu}_I} \right| \lambda \right\rangle \{\exp[-E_\lambda^0/(kT)] - \exp[-E_\eta^0/(kT)]\} \right).$$

and

$$\boldsymbol{\sigma}^{hf} = -\frac{1}{kT} \frac{\sum_{mn} Q_{mn} \left\langle n \left| \frac{\partial H^{(1)}}{\partial \mathbf{B}_0^T} \right| m \right\rangle \left\langle m \left| \frac{\partial H^{(1)}}{\partial \boldsymbol{\mu}_I} \right| n \right\rangle}{\sum_n \exp(-E_n^0/kT)}$$

with

$$Q_{mn} = \begin{cases} \exp[-E_n^0/(kT)] & \text{for } E_n^0 = E_m^0 \\ -\frac{kT}{E_m^0 - E_n^0} \{\exp[-E_m^0/(kT)] - \exp[-E_n^0/(kT)]\} & \text{for } E_n^0 \neq E_m^0 \end{cases}$$

**Appendix C: Orbital hyperfine coupling in the SE approach**

When *LS* coupling scheme applies, the PSO Hamiltonian describing dipolar interaction between nuclear spin and orbital angular momenta of unpaired electrons can be approximated with a Hamiltonian[42]

$$H^{IL} = 2 \frac{\mu_0}{4\pi} \frac{\hbar \gamma_I \mu_B}{r_I^3} \mathbf{L}_I^T \cdot \mathbf{I}$$

where $\mathbf{L}_I$ is the total electron orbital angular momentum at the position $\mathbf{r}_I$ defined relative to the NMR nucleus *I*. When the unpaired electrons are moving in a loop far away from the NMR nucleus, the dipolar magnetic field generated at the position of the nucleus can be approximately expanded as a function of the angular momentum $\mathbf{L}$ with respect to the paramagnetic center position as[43]

$$2 \frac{\mathbf{L}_I^T}{r_I^3} \approx 3 \mathbf{L}^T \cdot \frac{\mathbf{rr}^T}{r^5} - \frac{\mathbf{L}^T}{r^3}.$$

As a result,

$$H^{IL} \approx \frac{\mu_0}{4\pi} \frac{\hbar \gamma_I \mu_B}{r^3} \mathbf{L}^T \cdot \left( \frac{3\mathbf{rr}^T}{r^2} - \mathbf{1} \right) \cdot \mathbf{I}.$$

The orbital contribution $\mathbf{A}^L$ to the hyperfine coupling tensor (Eq. (42)) is then found by applying second-order perturbation theory with perturbation Hamiltonians $H^{IL}$ and $H^{LS} = \lambda \mathbf{S}^T \cdot \mathbf{L}$:

$$\mathbf{A}^L = \sum_m \frac{\left\langle 0 \left| \frac{\partial H^{IL}}{\partial \mathbf{I}^T} \right| m \right\rangle \left\langle m \left| \Sigma_i \frac{\partial H^{LS}}{\partial \mathbf{S}} \right| 0 \right\rangle + \left\langle 0 \left| \Sigma_i \frac{\partial H^{LS}}{\partial \mathbf{S}^T} \right| m \right\rangle \left\langle m \left| \frac{\partial H^{IL}}{\partial \mathbf{I}} \right| 0 \right\rangle}{E_0^0 - E_m^0}$$

$$= \frac{\mu_0}{4\pi} \frac{\hbar \gamma_I \mu_B}{r^3} \sum_m \frac{\left\langle 0 \left| \left(\frac{3\mathbf{rr}^T}{r^2} - \mathbf{1}\right) \cdot \mathbf{L} \right| m \right\rangle \langle m | \lambda \mathbf{L}^T | 0 \rangle + \langle 0 | \lambda \mathbf{L} | m \rangle \left\langle m \left| \mathbf{L}^T \cdot \left(\frac{3\mathbf{rr}^T}{r^2} - \mathbf{1}\right) \right| 0 \right\rangle}{E_0^0 - E_m^0}$$

$$= \frac{\mu_0}{4\pi} \frac{\hbar \gamma_I \mu_B}{r^3} 2\lambda \sum_m \frac{\langle 0|\mathbf{L}|m\rangle\langle m|\mathbf{L}^T|0\rangle}{E_0^0 - E_m^0} \cdot \left(\frac{3\mathbf{rr}^T}{r^2} - \mathbf{1}\right) = \frac{\mu_0}{4\pi} \frac{\hbar \gamma_I \mu_B}{r^3} (\mathbf{g} - g_e \mathbf{1}) \cdot \left(\frac{3\mathbf{rr}^T}{r^2} - \mathbf{1}\right)$$

where $\mathbf{g} = g_e \mathbf{1} + 2\lambda \sum_m \frac{\langle 0|\mathbf{L}|m\rangle\langle m|\mathbf{L}^T|0\rangle}{E_0^0 - E_m^0}$ is the definition of $\mathbf{g}$ tensor within the *LS* coupling scheme.

**Appendix D: Pseudocontact shift equations in limiting cases**

Let us consider the case of a $S = 1$ system with anisotropic $g$ values, axial ZFS tensor with $D > 0$, principal axes of the $\mathbf{g}$ and ZFS tensors coinciding, and electronically excited states with energy much larger than *kT*. The non-null elements of $\langle \mathbf{SS}^T \rangle$ are:

$$\langle S_x S_x \rangle = \langle S_y S_y \rangle = \frac{2kT}{D} \frac{1 - \exp\left(-\frac{D}{kT}\right)}{2\exp\left(-\frac{D}{kT}\right) + 1}, \quad \langle S_z S_z \rangle = \frac{2\exp\left(-\frac{D}{kT}\right)}{2\exp\left(-\frac{D}{kT}\right) + 1}$$

and thus, in the principal frame of the $\mathbf{g}$ tensor (see Eq. (6)),

$$\langle \mathbf{S}_C \rangle = -\frac{\mu_B}{kT} \begin{pmatrix} g_{xx} \frac{2kT}{D} \frac{1-\exp\left(-\frac{D}{kT}\right)}{2\exp\left(-\frac{D}{kT}\right)+1} B_x \\ g_{yy} \frac{2kT}{D} \frac{1-\exp\left(-\frac{D}{kT}\right)}{2\exp\left(-\frac{D}{kT}\right)+1} B_y \\ g_{zz} \frac{2\exp\left(-\frac{D}{kT}\right)}{2\exp\left(-\frac{D}{kT}\right)+1} B_z \end{pmatrix}$$

and (see Eq. (12))

$$\chi = \frac{\mu_0 \mu_B^2}{kT} \begin{pmatrix} g_{xx}^2 \frac{2kT}{D} \frac{1-\exp\left(-\frac{D}{kT}\right)}{2\exp\left(-\frac{D}{kT}\right)+1} & 0 & 0 \\ 0 & g_{yy}^2 \frac{2kT}{D} \frac{1-\exp\left(-\frac{D}{kT}\right)}{2\exp\left(-\frac{D}{kT}\right)+1} & 0 \\ 0 & 0 & g_{zz}^2 \frac{2\exp\left(-\frac{D}{kT}\right)}{2\exp\left(-\frac{D}{kT}\right)+1} \end{pmatrix}.$$

Therefore, using Eqs. (34) and (38) of the SE approach, to the first order in $D/(kT)$,

$$\delta^{pc} = \frac{\mu_0}{4\pi} \frac{\mu_B^2}{9k} \left[ (2g_{zz}^2 - g_{xx}^2 - g_{yy}^2) \frac{3\cos^2\theta - 1}{r^3} \left(1 - \frac{D}{6kT} \frac{4g_{zz}^2 + g_{xx}^2 + g_{yy}^2}{2g_{zz}^2 - g_{xx}^2 - g_{yy}^2}\right) + (g_{xx}^2 - g_{yy}^2) \frac{3\sin^2\theta \cos 2\varphi}{r^3} \left(1 + \frac{D}{6kT}\right) \right]$$

,

in agreement with Kurland and McGarvey.[3] Analogous equations are obtained for other spin multiplicities.[3]

In comparison, the non-symmetric tensor $\chi'$ of the QC approach for the same $S = 1$ system is according to Eq. (22)

$$\chi' = \frac{\mu_0 \mu_B^2 g_e}{kT} \begin{pmatrix} g_{xx} \frac{2k}{D} \frac{1-e\left(-\frac{D}{kT}\right)}{1+2\text{ex}\left(-\frac{D}{kT}\right)} & 0 & 0 \\ 0 & g_{yy} \frac{2kT}{D} \frac{1-\exp\left(-\frac{D}{kT}\right)}{1+2\text{ex}\left(-\frac{D}{kT}\right)} & 0 \\ 0 & 0 & g_{zz} \frac{2\exp\left(-\frac{D}{kT}\right)}{1+2e\left(-\frac{D}{kT}\right)} \end{pmatrix}$$

and, using Eq. (23), to the first order in $D/(kT)$ we obtain

$$\delta^{pc} = \frac{\mu_0}{4\pi} \frac{\mu_B^2 g_e}{9kT} \left[ (2g_{zz} - g_{xx} - g_{yy}) \frac{3\cos^2\theta - 1}{r^3} \left(1 - \frac{D}{6kT} \frac{4g_{zz} + g_{xx} + g_{yy}}{2g_{zz} - g_{xx} - g_{yy}}\right) \right.$$

$$\left. + (g_{xx} - g_{yy}) \frac{3\sin^2\theta \cos 2\varphi}{r^3} \left(1 + \frac{D}{6kT}\right) \right]$$

where $2g_{zz} - g_{xx} - g_{yy} = 2\Delta g$ is twice the g-anisotropy and $g_{xx} - g_{yy}$ is the g-asymmetry. Essentially, all **g** tensor components that appeared above in the SE approach in the second power are here in the QC approach in the first power, correspondingly to the transition from $\chi$ and Eq. (34) for dipolar shielding to $\chi'$ and Eq. (23). We reiterate that in the QC framework formulated within the point-dipole approximation, the spin-orbital contribution to the hyperfine shift is neglected.[31]

**Figure 1.** A schematic comparison of SE and QC frameworks for PNMR shift theory. In the SE picture (left), the nuclear spin, total electron spin and total electron orbital angular momenta **I**, **S**, and **L** are considered together with contributions to the total energy from the interaction between all pairs of the associated magnetic moments. In the QC framework (right), interactions are considered on the level of single particles, *i.e.* between all pairs of angular momenta **I**, $\mathbf{s}_i$, and $\mathbf{l}_i$, as well as the Zeeman interactions between these angular momenta and the external magnetic field $\mathbf{B}_0$ (not shown)

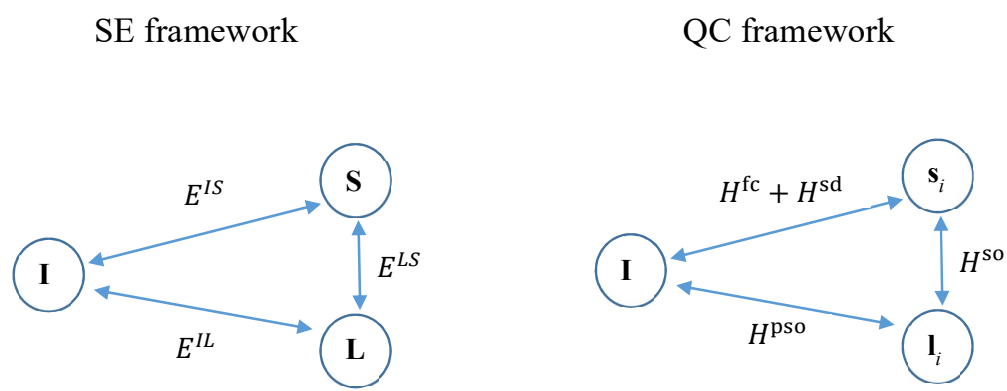

**Figure 2.** The average electron spin magnetic moment ⟨**μ**_S⟩ results from the different population of the electron spin states and is directed along the magnetic field direction. The average total electron magnetic moment (comprising spin and orbital contributions) ⟨**μ**⟩ is not parallel to ⟨**μ**_S⟩ in the presence of *g*-anisotropy.

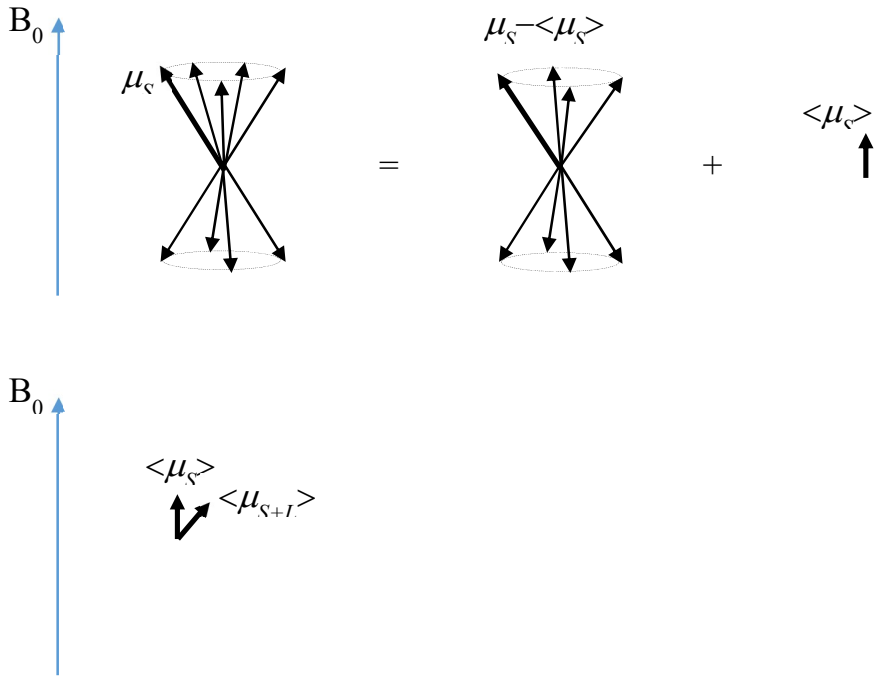

**Figure 3.** Summary of the main equations in the derivation of the pseudocontact shift and magnetic susceptibility tensor in the SE and QC approaches.

## SE framework[1]

$$E^{\text{dip}} = -\frac{\mu_0}{4\pi r^3} \langle \boldsymbol{\mu}_{S+L} \rangle^T \cdot \left(\frac{3\mathbf{rr}^T}{r^2} - 1\right) \cdot \boldsymbol{\mu}_I$$

$$\sigma^{\text{dip}} \approx -\frac{\mu_0}{4\pi r^3} \frac{\mu_B^2}{kT} \mathbf{g} \cdot \langle \mathbf{SS}^T \rangle \cdot \mathbf{g}^T \cdot \left(\frac{3\mathbf{rr}^T}{r^2} - 1\right)$$

$$\boldsymbol{\chi} = \frac{\mu_0 \mu_B^2}{kT} \mathbf{g} \cdot \langle \mathbf{SS}^T \rangle \cdot \mathbf{g}^T$$

$$\delta^{\text{pc}} = \delta(\boldsymbol{\chi})$$

## QC framework[2]

$$E^{\text{dip}} \approx -\frac{\mu_0}{4\pi r^3} \boldsymbol{\mu}_S^T \cdot \left(\frac{3\mathbf{rr}^T}{r^2} - 1\right) \cdot \boldsymbol{\mu}_I$$

$$\sigma^{\text{dip}} \approx -\frac{\mu_0}{4\pi r^3} \frac{\mu_B^2}{kT} g_e \mathbf{g} \cdot \langle \mathbf{SS}^T \rangle \cdot \left(\frac{3\mathbf{rr}^T}{r^2} - \mathbf{1}\right)$$

$$\boldsymbol{\chi}' = \frac{\mu_0 \mu_B^2}{kT} g_e \mathbf{g} \cdot \langle \mathbf{SS}^T \rangle$$

$$\delta^{\text{pc}} = \delta(\boldsymbol{\chi}')$$

1. Kurland & McGarvey, J.Magn.Reson. 2, 286–301 (1970)

2. Vaara et al. J. Chem. Theory Comput. 11, 4840–4849 (2015)
Benda et al., Angew. Chem. Int. Ed. 55, 14713–14717 (2016)

# Pseudocontact shifts and paramagnetic susceptibility in traditional and quantum chemistry theories

Giacomo Parigi,[1,2,a]
Ladislav Benda,[3,a]
Enrico Ravera,[1,2]
Maurizio Romanelli,[4]
and Claudio Luchinat[1,2]

[1]*Magnetic Resonance Center (CERM) and Interuniversity Consortium for Magnetic Resonance of Metallo Proteins (CIRMMP), Via L. Sacconi 6, 50019 Sesto Fiorentino, Italy*
[2]*Department of Chemistry "Ugo Schiff", University of Florence, Via della Lastruccia 3, 50019 Sesto Fiorentino, Italy*
[3]*Centre de RMN à Très Hauts Champs, FRE 2034 CNRS / ENS de Lyon / UCB Lyon 1, 5 rue de la Doua, 69100 Villeurbanne (Lyon), France*
[4]*Department of Earth Sciences, University of Florence, via G. La Pira 4, 50121 Florence, Italy*

**Contents**

# Hyperfine shielding terms in QC and SE frameworks

**Table S1.** Comparison of $^{13}$C hyperfine shielding terms (in ppm) arising in QC and SE theory frameworks for the atoms close to the Co$^{II}$ center of the CoMMP-12 protein.[a]

| PDB atom identifiers | | | theory framework: shielding term: HFC calc. method: | QC, SE $\sigma^{con}$ [c] PBE0 | QC $\sigma^{sd}$ [d] PBE0 | QC $\sigma^{so}$ [e] PBE | SE $\sigma^{dip-LS}$ [f] PDA [i] | QC, SE $\sigma^{sd}$ [g] PDA [i] | SE $\sigma^{L}$ [h] PDA [i] |
|---|---|---|---|---|---|---|---|---|---|
| Atom type | Residue type | no. | $r_{M,C}$ [b] | | | | | | |
| CA [j] | VAL | 217 | 10.0 | -0.3 | -2.2 | -0.1 | -3.7 | -2.2 | -1.5 |
| C' | VAL | 217 | 8.6 | -1.6 | -3.9 | -0.2 | -6.5 | -3.9 | -2.6 |
| CA | HIS | 218 | 6.2 | -30.9 | -9.3 | -0.5 | -14.3 | -8.6 | -5.7 |
| C' | HIS | 218 | 5.7 | 0.1 | -14.6 | -0.6 | -25.0 | -14.9 | -10.1 |
| CB | HIS | 218 | 5.6 | -27.9 | -6.9 | -0.2 | -10.4 | -6.4 | -4.0 |
| CG | HIS | 218 | 4.2 | -584.9 | -15.2 | -2.7 | -18.7 | -11.9 | -6.8 |
| CD2 | HIS | 218 | 3.0 | 268.6 | -44.1 | -2.7 | -69.2 | -41.9 | -27.3 |
| CE1 | HIS | 218 | 3.1 | -1009.4 | 52.6 | 6.8 | 73.1 | 41.1 | 32.0 |
| CA | GLU | 219 | 6.2 | 0.1 | -9.5 | -0.4 | -16.8 | -9.7 | -7.2 |
| C' | GLU | 219 | 7.2 | 0.1 | -7.8 | -0.4 | -13.6 | -7.9 | -5.6 |
| CB | GLU | 219 | 6.8 | 0.4 | -3.2 | -0.2 | -6.0 | -3.2 | -2.8 |
| CG | GLU | 219 | 6.3 | -5.9 | 0.8 | 0.0 | 0.6 | 1.0 | -0.4 |
| CD | GLU | 219 | 5.0 | 6.1 | 5.4 | 0.2 | 8.0 | 6.2 | 1.9 |
| CA [j] | ILE | 220 | 9.4 | 0.0 | -4.3 | -0.2 | -7.3 | -4.3 | -3.0 |
| C' | ILE | 220 | 9.1 | 0.0 | -5.2 | -0.2 | -8.7 | -5.2 | -3.5 |
| CA | GLY | 221 | 8.3 | 0.1 | -6.6 | -0.3 | -10.9 | -6.6 | -4.3 |
| C' | GLY | 221 | 7.8 | -0.8 | -8.0 | -0.3 | -13.0 | -7.9 | -5.1 |
| CA | HIS | 222 | 6.7 | -67.3 | -12.6 | -0.6 | -17.5 | -10.7 | -6.8 |
| C' | HIS | 222 | 8.0 | -16.5 | -5.7 | 0.3 | -9.2 | -5.6 | -3.6 |
| CB | HIS | 222 | 5.6 | -21.2 | -16.7 | -0.5 | -25.0 | -15.2 | -9.8 |
| CG | HIS | 222 | 4.2 | -679.7 | -39.7 | -3.3 | -58.9 | -36.0 | -22.9 |
| CD2 | HIS | 222 | 3.1 | 30.9 | -94.6 | -7.1 | -144.9 | -87.2 | -57.7 |
| CE1 | HIS | 222 | 3.1 | -943.1 | -6.8 | 13.3 | -58.4 | -38.3 | -20.1 |
| CA [j] | SER | 223 | 9.9 | -0.7 | -2.9 | -0.4 | -4.7 | -2.8 | -1.8 |
| CA [j] | LEU | 226 | 8.8 | 0.0 | -2.2 | -0.1 | -3.4 | -2.2 | -1.2 |
| C' | LEU | 226 | 7.6 | -3.5 | -1.8 | -0.1 | -2.9 | -1.9 | -1.0 |
| CA | GLY | 227 | 6.9 | -0.3 | 3.4 | 0.1 | 5.5 | 3.3 | 2.3 |
| C' | GLY | 227 | 6.4 | -0.9 | 7.6 | 0.3 | 12.5 | 7.6 | 4.9 |
| CA | HIS | 228 | 6.0 | -10.1 | 13.0 | 0.6 | 21.7 | 13.5 | 8.2 |
| C' | HIS | 228 | 7.4 | 0.7 | 7.2 | 0.3 | 11.7 | 7.2 | 4.4 |
| CB | HIS | 228 | 5.6 | -39.4 | 17.4 | 0.7 | 24.1 | 15.3 | 8.8 |
| CG | HIS | 228 | 4.2 | -570.6 | 32.1 | -1.3 | 49.7 | 31.9 | 17.8 |
| CD2 | HIS | 228 | 3.1 | -17.3 | 99.6 | 5.0 | 162.0 | 101.7 | 60.3 |
| CE1 | HIS | 228 | 3.0 | -944.6 | -36.9 | -3.2 | -38.0 | -17.4 | -20.5 |
| CA [j] | SER | 229 | 9.1 | 0.0 | 3.9 | 0.2 | 6.3 | 3.9 | 2.4 |
| C1 | NGH | 269 | 5.6 | 1.7 | -1.0 | 0.0 | -0.5 | -1.0 | 0.4 |
| C2 | NGH | 269 | 5.7 | -1.6 | 3.0 | 0.1 | 6.1 | 3.0 | 3.1 |
| C3 | NGH | 269 | 5.2 | 1.0 | 5.6 | 0.3 | 10.8 | 5.8 | 5.0 |
| C4 | NGH | 269 | 4.9 | -1.3 | 9.2 | 0.4 | 16.0 | 9.2 | 6.8 |
| C5 | NGH | 269 | 4.8 | -1.4 | 9.5 | 0.5 | 16.9 | 9.8 | 7.2 |
| C6 | NGH | 269 | 5.0 | -2.0 | 0.7 | 0.4 | 1.9 | 0.5 | 1.4 |
| C7 | NGH | 269 | 6.9 | 0.4 | 3.2 | 0.1 | 6.0 | 3.3 | 2.8 |
| C9 | NGH | 269 | 5.0 | -32.6 | -25.7 | -1.2 | -41.0 | -25.1 | -15.9 |
| C10 | NGH | 269 | 4.4 | -552.6 | -16.9 | -0.3 | -17.0 | -10.9 | -6.2 |
| C11 | NGH | 269 | 2.9 | 824.3 | -6.6 | 1.4 | -16.7 | -11.7 | -5.0 |
| C12 | NGH | 269 | 5.9 | -10.1 | -16.0 | -0.7 | -25.4 | -15.4 | -10.0 |
| C13 | NGH | 269 | 7.3 | -2.3 | -8.0 | -0.4 | -12.9 | -7.9 | -5.0 |
| C14 | NGH | 269 | 6.1 | -0.3 | -14.9 | -0.7 | -24.9 | -14.9 | -9.9 |

[a]All data were calculated using the model structure of the CoMMP-12 metal center from ref. [1], optimized at the DFT PBE0-D3BJ level. All hyperfine shielding calculations utilized the *g*- and ZFS tensors obtained at the NEVPT2 level of theory as described in ref. [1]. [b]Distance from the metal in the PDB structure 1RMZ (in Å). [c]Isotropic hyperfine shielding due to Fermi-contact hyperfine coupling was taken from ref. [1], corresponding to Eqs. (16) and (19) of the main text, given here for context. [d]Isotropic hyperfine shielding due to spin-dipolar hyperfine coupling was taken from ref. [1], corresponding to Eqs. (16) and (18) of the main text. [e]Isotropic hyperfine shielding due to spin-orbital hyperfine coupling calculated according to Eq. (24). [f]Isotropic hyperfine shielding due to the dipolar

hyperfine coupling in the *LS* coupling approximation of the SE framework including both the spin-dipolar and the "orbital" term, calculated according to Eqs. (39) and (41). [g]Isotropic hyperfine shielding due to spin-dipolar hyperfine coupling within the point-dipole approximation was taken from ref. [1], corresponding to Eq. (21). [h]Isotropic hyperfine shielding due to "orbital" hyperfine coupling in the *LS* coupling approximation of the SE framework, calculated according to Eqs. (39) and (42). [i]Point-dipole approximation. [j]Terminal carbon atom in the model structure.

## Computational details

The spin-orbital term $\mathbf{A}^{\text{so}}$ of the hyperfine coupling tensor was calculated on a DFT level employing PBE functional[2] and IGLO-II basis[3] as implemented in ORCA program.[4]